\documentclass[12pt]{article}
\usepackage{amsfonts,amsmath,amsthm}
\usepackage{graphicx}

\def\be{\begin{equation}}
\def\ee{\end{equation}}
\def\ba{\begin{array}}
\def\ea{\end{array}}

\def\bee{\begin{eqnarray}}
\def\eee{\end{eqnarray}}

\def\dis{\displaystyle}
\def\cR{{\cal R}}

\begin{document}
\title {SARS outbreaks in Ontario, Hong Kong and Singapore: the role
   of diagnosis and isolation as a control mechanism}

\author{G.\ Chowell$^{1,2}$, P.\ W.\ Fenimore$^{1}$, M.\ A.\
Castillo-Garsow$^{3}$, and C.\ Castillo-Chavez$^{1,2}$\\
\footnotesize $^{1}$ Center for Nonlinear Studies \\
\footnotesize MS B258 \\
\footnotesize Los Alamos National Laboratory \\
\footnotesize Los Alamos, NM 87545 \\
\footnotesize $^{2}$ Department of Biological Statistics and
Computational Biology\\
\footnotesize Cornell University \\
\footnotesize Warren Hall, Ithaca, NY 14853\\
\footnotesize $^{3}$ Universidad de Colima \\
\footnotesize Facultad de Ciencias \& Facultad de Letras\\
\footnotesize Col. Villas de San Sebasti\'an \\
\footnotesize 28045, Colima, Colima. M\'exico. \\
\footnotesize LA-UR-03-2653 \\
}

\date{}

\maketitle

\begin{abstract}
In this article we use global and regional data from the SARS epidemic
in conjunction with a model of susceptible, exposed, infective,
diagnosed, and recovered classes of people (``SEIJR'') to extract
average properties and rate constants for those populations. The model
is fitted to data from the Ontario (Toronto) in Canada, Hong Kong in 
China and
Singapore outbreaks
and predictions are made based on various assumptions and
observations, including the
current effect of isolating individuals diagnosed with SARS. The
epidemic
dynamics for Hong Kong and Singapore appear to be different from the
dynamics in Toronto, Ontario. Toronto shows a very rapid
increase in the number of cases between March 31st and April 6th,
followed by a {\it significant} slowing in the number of new cases. 
We explain
this as the result of an increase in the diagnostic rate and in the 
effectiveness of
patient isolation after March 26th. Our best estimates are consistent
with SARS eventually being contained in Toronto, although the time of
containment is sensitive to the parameters in our model. It is shown
that despite the empirically modeled heterogeneity in transmission, 
SARS'
average reproductive number is $1.2$, a value quite similar to that
computed for some strains of influenza \cite{CC2}.
Although it would not be surprising to see
levels of SARS infection higher than ten per cent in some regions of
the world (if unchecked), lack of data and the observed heterogeneity 
and sensitivity of
parameters prevent us from predicting the long-term impact of SARS. 
The
possibility that 10 or more percent of the world population 
\textit{at risk} could
eventually be infected with the virus in conjunction with a mortality
rate of three-to-seven percent or more, and indications of {\it 
significant}
improvement in Toronto support the stringent measures that
have been taken to isolate diagnosed cases.
\end{abstract}

\section{Introduction}

Severe acute respiratory syndrome (SARS) is a new respiratory disease
which was
first identified in China's southern province of Guangdong. SARS is
not merely a local endemic disease: it poses a serious risk to the
medical community, is a threat to international travelers, is
having a substantial negative economic impact in parts of East Asia
and is spreading world-wide.
The serious danger SARS poses to the medical community is illustrated
by the numerous cases of transmission to health-care
workers. Startlingly, the man who awakened the world to the dangers of
SARS, Dr.\ Carlo Urbani, succumbed to the disease. Cases of
transmission between aircraft passengers are suspected, and relatively
short visits to epidemic regions have resulted in infection. The most
striking feature of SARS, however, has proven to be its ability to
rapidly spread on a global scale. One man with SARS made $7$ flights:
from Hong Kong to M\"unich to Barcelona to Frankfurt to London, back
to M\"unich and Frankfurt before finally returning to Hong Kong
\cite{NYT4}. Another individual, a $26$-year-old airport worker,
appears to have transmitted the disease to $112$ people
\cite{NYT1}. Clearly, there is an unfortunate interaction between the
incubation period of the virus, the widely distributed severity and
infectiousness of SARS in different people and the speed and volume
of passenger air
travel. The adverse economic impact in parts of East Asia far exceeds
the disruption of previous outbreaks of avian influenza, earning
comparison with the 1998 financial market crisis in that part of the
world \cite{MSNBC1,MSNBC2,MSNBC3}. Although the causative agent of
SARS has been determined \cite{Drosten,Ksiazek}, a detailed
understanding of the causative virus' pathogenticity and routes of
transmission and the dynamics of the epidemic is still at a very
early stage.  It is uncertain how the virus is transmitted: by droplet
or airborne transmission or person-to-person contact. The recent
development of laboratory tests promises to improve the
epidemiological situation somewhat \cite{faster}.

SARS is a public health crisis on a scale rarely seen. The obvious
question in such a crisis is, ``can SARS be contained?'' In this
study, we
report transmission parameters and epidemic dynamics from a
model based on classes of people who are susceptible, exposed,
infectious, diagnosed, and recovered
(``SEIJR'') that includes the effect of patient isolation. Our model
is consistent with the possibility of containment in Toronto, Ontario.

\section {SARS epidemiology and related issues}
SARS was first identified in November 2002 in the Guongdong Province
of China \cite{WP1}. By February $26$, $2003$ officials in Hong Kong
reported their first cases of SARS and no later than March
$14^{\mathrm{th}}$ of this year the virus reached Canada
\cite{WP11}. As of April $17^{\rm th}$, Canada is the only location
outside of Asia which has seen deaths as a result of SARS ($13$ so
far) \cite{MSNBC5}.  U.\,S.\ health officials are currently
investigating $199$ cases in $34$ states (Apri 17, 2003) \cite{WP10}.

An individual exposed to SARS may become infectious after an
incubation period of $2-7$ days (or longer) \cite{CDC4} with $3-5$
days being most common \cite{FIRECHIEF1}. Most infected individuals 
either
recover, typically after $7$ to $10$ days,
or suffer $4\%$ mortality or higher \cite{CHEALTH1, promed2,BBC1}.
SARS appears to be most serious in people over age $40$, especially
those who have other medical problems such as heart or liver disease.
Its symptoms are similar to pneumonia or other respiratory ailments
and include a high fever ($\ge 38^{\circ}$\,C), shortness of
breath,
dry cough, headache, stiff or achy muscles, fatigue and diarrhea
\cite{WP12}. These symptoms, however, are not uniform. In the US, for
example, the disease seems to be a milder one than in Asia
\cite{MSNBC6}. The result has been that SARS was, and for the moment
remains, a diagnosis of exclusion.

Presently, there is no treatment for SARS \cite{NEWSWEEK1} and
diagnostic tests are just becoming available \cite{faster}. The 
mortality rate is reported
to be $4$\% or higher world-wide\cite{promed2, BBC1}. Experts 
estimate that between
$80$ and
$90$ percent of people with SARS recover without medical intervention,
while the condition of the remaining victims requires medical care
\cite{WP12}. As of April $17$, $2003$, the World Health Organization
(WHO)
reported $3,389$ cases (a mixture of probable or suspected cases) in
$26$ countries. $165$ victims are reported to have died \cite{WP10}.

Although researchers in the Erasmus Medical Center in Rotterdam
recently demonstrated that a coronavirus (some of which produce
common colds) is the causative agent of SARS, the mode of transmission
still remains unknown  \cite{WP10}. The current hypothesis is that
SARS is transmitted mainly by close person-to-person contact which may
explain the relatively slow transmission scale. However, it could 
also be
transmitted through contaminated objects, air or by other unknown
ways \cite{CDC1}. It is also a mystery how the disease originated,
whether in
birds, pigs or other animals, nor is it known if the
origin is rural or urban \cite{NYT3}.

In this article, a simple model for SARS outbreaks is formulated (see
\cite{BC}). The model is used in conjunction with {\em global and
local}\,\, SARS data to estimate the initial growth rate of the SARS
epidemic. These rates are used to estimate SARS' {\em basic
reproductive number}, $R_0$, the classical epidemiological measure
associated with the reproductive power of a disease. $R_0$ estimates
the {\em average}{} number of secondary cases of infection generated
by a {\em typical} infectious individual in a population of
susceptibles \cite{Diekmann1} and hence, it is used to
estimate the initial growth of a SARS outbreak.
We estimate (using data from Ontario,
Hong Kong and Singapore) that $R_0$ is about
$1.2$. This value is not too different from past estimates of $R_0$ 
for influenza (see
\cite{CC2}) despite the fact that {\em superspreaders} of SARS have
been identified. In fact, the parameter values resulting on this 
$R_0$, on our
population-scaled model, can lead to extremely high levels of 
infection).
We show, via simple extrapolation, that the estimated rate
of growth is consistent with the reported date for the first cases of
SARS in Hong Kong, however the first cases in Toronto may be
several weeks earlier than the February 23 date
of the first case reported by the Canadian Health Ministries
\cite{CMH}. Our best
``rough'' estimate for Toronto is that the first case occurred 
sometime around
January 29th, and not later than February 28th.  The data for Hong
Kong are fitted by fixing the parameters $k$, $\delta$ and
$\gamma_1$ based on estimates of the observed rates for the
corresponding processes. The growth rate $\beta$ is estimated from
observed ``model-free'' exponential growth in Singapore and Hong
Hong. The {\em average} diagnostic rate $\alpha$ and
the measure of heterogeneity between the two susceptible classes $p$
and the effectiveness of patient isolation measures (related to $l$)
are then varied to fit the initial data for Hong Kong and Singapore. 
To model
the data in Toronto, we must postulate that the parameters describing
the rate of
diagnosis ($\alpha$) and isolation ($l$) in the Canadian outbreak
changed radically on
March $27$. Two hospitals in Toronto were closed about that time:
Scarborough Grace Hospital on March $25^{\mathrm{th}}$ and York
Central Hospital on March $28^{\mathrm{th}}$ \cite{closure}. The
remainder of this
article is organized as follows: Section $4$ introduces the basic
model and gives a formula for the basic reproductive number $R_0$; Section $5$
describes the results of simulations and connections to data; and,
Section $6$ collects our final thoughts.

\section{SARS' Transmission Model}
U.\,S.\ data is limited and sparsely distributed \cite{promed1,
reporter1}
while the quality of China's data is hard to evaluate
\cite{underreporting}. On the other hand, there appears to be enough
data for Toronto \cite{CMH}, Singapore and Hong Kong \cite{WHO} to
make limited
preliminary predictions using a model that includes the effects of
\textit{suspected} mechanisms for the spread of SARS. Limited data and
inconclusive epidemiological information place severe restrictions on
efforts to model the global spread of the SARS etiological agent.

Thus, we model {\em single} outbreaks, ignoring demographic processes
other than the impact of SARS on survival. The model is applied to
data from Toronto, Hong Kong and Singapore. Because the
outbreak dynamics in Singapore and Hong Kong are different from those
in Toronto, some of the results may only be indicative of what is
happening in those regions of the world (in particular our parameters
$\alpha$ and $l$ may change). The situation must be re-evaluated
frequently as
SARS continues its travels around the world.

Here we describe a model that incorporates, in a \textit{rather crude}
way, some of the important characteristics suggested in the literature
(unequal susceptibility, symptomatic and asymptomatic individuals,
mode of transmission, superspreaders, etc.)
\cite{CDC1,CDC2,CDC3,promed1}. The goal is to use the results for
single outbreaks as a first step in our efforts to gauge the {\it
global} impact of SARS. Hence, we focus on three ``closed''
populations
(Southern Ontario (Toronto), Singapore and Hong Kong) and postulate 
differences
in the degree of susceptibility to SARS \cite{NYT1,WP12}. These
differences may be due to variations in contact rates, age-dependent
susceptibility or ``unknown'' genetic factors. This last
assumption is handled (in a rather crude and arbitrary way) via the 
introduction of two distinct susceptible
classes: $S_1$, the most susceptible, and $S_2$, less so. Initially,
$S_1 = \rho N$ and $S_2 = (1-\rho) N$ where $\rho$ is the proportion
of the population size $N$ that is initially at higher risk of SARS
infection. The
parameter $p$ is a measure of reduced susceptibility to SARS
in class $S_2$ \cite{WP12, NYT1}.  $E$ (``exposed'') denotes the class
composed of asymptomatic, possibly infectious (at least some of the 
time)
individuals. Typically, it takes some time before asymptomatic
infected individuals become infectious.  The possibility of
limited transmission from class $E$ is included,
in a rather crude way, via the parameter $q$ (see Table $1$). The
class $I$ denotes infected, symptomatic, infectious, and undiagnosed
individuals.  $I$-individuals move into the diagnosed class $J$ at the
rate $\alpha$.  Individuals recover at the rates $\gamma_1$ ($I$
class) and $\gamma_2$ ($J$ class). The rate $\delta$ denotes SARS'
disease-induced mortality. The classes $R$ is included to
keep track of the cumulative number of diagnosed and recovered,
respectively. Furthermore, it is
assumed that diagnosed individuals
are handled with {\it care}. Hence, they might not be ({\em
effectively})
as infectious as those who have not been diagnosed (if $l$ is
small). The parameter $l$ takes into account their reduced impact on 
the transmission process
(small $l$ represents effective measures taken to isolate diagnosed
cases and {\it visa versa}). Table $1$ includes parameters'
definitions and the initial values used.  Our SARS epidemiological
model
is given by the following nonlinear system of differential equations:

\begin{equation}
\label{eq1}
\begin{array}{rcl}
      {\dis \dot{S_1}}&=&-\beta S_1 \frac{(I+qE+lJ)}{N},\\
      {\dis \dot{S_2}}&=&-\beta p  S_2 \frac{(I+qE+lJ)}{N},\\
      {\dis \dot{E}}&=&\beta(S_1+ pS_2) \frac{(I+qE+lJ)}{N} - kE,\\
      {\dis \dot{I}}&=& kE - (\alpha+\gamma_1 + \delta)I,\\
      {\dis \dot{J}}&=&\alpha I - (\gamma_2+\delta)J,\\
      {\dis \dot{R}}&=&\gamma_1 I + \gamma_2 J,\\
\end{array}
\end{equation}
which is refered to as ``SEIJR,'' after the variables used to name the
classes.
\\

The values of $p$ and $q$ are not known and are fixed arbitrarily while $l$ and $\alpha$ are
varied and optimized to fit the existing data (least-squares criterion) for Hong Kong,
Singapore and Toronto. We did not explored the sensitivity of the model to variations 
in $p$ and $q$ because they are not known and cannot be controlled. All other parameters were roughly 
estimated from data \cite{CMH, WHO} and current literature 
\cite{CDC1,CDC4, FIRECHIEF1,CHEALTH1}. In particular, the 
transmission 
rate $\beta$ is calculated from the
dominant root of the third order equation obtained from the 
linearization
around the disease-free equilibrium \cite{Diekmann1}. The
parameters $l$ and $\alpha$ were allowed to vary when fitting the data
for each location (Singapore, Hong Kong and Toronto). Some
restrictions apply, for example, the value of $\alpha >
\gamma_1$. We also require that $1/\gamma_2 = 1/\gamma_1 - 1/\alpha$,
a statement that members of the diagnosed class $J$ recover at the 
same
rate as members of the undiagnosed class $I$. $1/\gamma_1$ has been
reported to be between $7$ and $10$ days \cite{promed2, CHEALTH1}.
\noindent From the second generator approach \cite{Diekmann1}, we 
obtain the following expression for the basic reproductive number:

\begin{equation}
\label{eqn2}
\begin{array}{rcl}
  {\cR_0} & = & \left\{\beta \left[\rho+p(1-\rho)\right]\right\}
  \left\{\frac{q}{k} + \frac{1}{\alpha + \gamma_1 + \delta} +
  \frac{\alpha l}{(\alpha + \gamma_1 + \delta)(\gamma_2 + 
\delta)}\right\}\\
\end{array}
\end{equation}

\noindent which can be easily given an epidemiological
  interpretation. The use
of parameters estimated from Hong Kong (Table $1$) gives a
values of $R_0 = 1.2$ (Hong Kong) and $R_0=1.2$ (Toronto, assuming
exponential growth) and $R_0=1.1$ (Singapore).

\section{Simulation Results}

Initial rates of growth for SARS outbreaks in different parts of the
world (see Figure \ref{myfig00}) are computed using the data provided
by WHO \cite{WHO} and the Canadian Ministry of Health
\cite{CMH}. These rates are computed
exclusively from the number of cases reported between March $31$ and
April $14$. The values obtained are $0.0405$ (world data), $0.0496$
(Hong Kong), $0.054$ (Canada), $0.054$ (Toronto) and $0.037$
(Singapore).

For our numerical simulations, we start with an infectious individual
(not yet diagnosed, $I(0)=1$) and \textit{crude} estimates for the 
start of SARS outbreaks
($t_0$) are obtained from the formula $ t_0 =
t - (\frac{1}{r} \log (x(t)))$, which assumes initial exponential
growth ($r$, the estimated ``model-free'' rate of growth from the time
series $x(t)$ of the cumulative number of SARS cases).
Results for Toronto, Hong Kong, Singapore and aggregated
world data are shown in Table
$2$. The estimated ``world'' start of the outbreak is
November $5$, a date consistent with the
fact that the first SARS case was detected in Guangdong, China in
November \cite{WP1}. These dates are used as the starting time of the
respective outbreaks. \\

For the case of the Province of Ontario, Canada the total
population $N$ is approximately $12$ million. We assume that the 
population
at \textit{major} risk of SARS infection lives in Ontario's southern 
part 
(particularly Toronto), and is
approximately $40\%$ of the total population ($\rho=0.4$ in our
model). It is worth pointing out that this value of $\rho$ is not
critical (that is, the most sensitive) in the model. The
``model-free'' approximately exponential growth rates for the various 
regions of the world are roughly similar \emph{except for
Canada} from March 31st (day 61) to April 6th (beginning the
day of the jump in the number of reported Canadian cases), the
number of diagnosed cases grew $\sim\exp(0.081 t)$, where $t$ is
measured in days. This rate is substantially higher than elsewhere in
the world. In the subsequent week (beginning April 7th, day 68) the
number of probable or suspected
Canadian cases rapidly rolls over to a smaller growth
rate not too far from the rest of the world. We conclude, based on
the coincidence of the Canadian hospital closures, the jump in the
reported number of Canadian SARS cases on March 31st and the rapid
rise in recognized cases in the following week, that Canadian doctors 
were
rapidly diagnosing pre-existing cases of SARS (in either class $E$ or 
$I$ on March
26th). If we make the assumption that the fundamental
disease spreading parameters other than $\alpha$ and $l$ are roughly
constant throughout the world prior to March 26th, we can reach two
important
conclusions. Beginning on March 26th, in Toronto:
\begin{itemize}
\item $\alpha$ changed from a
number $1/\alpha \approx 1/\gamma_1 - 2 \approx 6$ days to
$1/\alpha_1 \leq 3$ days, and
\item $l$ changed from an uncertain and
relatively large value $l> 1/2$ to $l\leq 0.1$.
\end{itemize}
If we assume that the fundamental growth rate $\beta$ is essentially 
constant from one region of the world to another, it is difficult for 
our model to produce growth rates $r$ well above the world average,
except as a transient response to differences in diagnostic rate
$\alpha$ (due to delays in response or change in policy). Similarly, 
the SEIJR model requires fairly small values of
$l$ to achieve a rapid roll-over in the growth rate of recognized
cases. The parametric details of how a ``second'' initial condition
for Toronto on March 26th is generated
do not affect the qualitative aspect of this argument: the Canadian
data prior to March 31st
(the day of the large jump) are probably not as meaningful as data
after
that date, and hence only bound the model from below prior to March
26th. The essential aspect of this before-and-after hospital closure
argument is that there were substantially more undiagnosed people in
classes $E$ and $I$ than in class $J$ on March 26th. This is a
reasonable assumption
given that the number of cases reported by Canadian officials more
than
double from March 30th to March 31st.
The introduction of behavioral changes starting on March $26$ ($t=57$
days),alters the fate of the disease in a dramatic
fashion (see Table $3$).\\

Fitting the model to the Hong Kong and Singapore data is carried out
in a similar
fashion with $\rho = 0.4$, (Hong Kong has about $7.5$ million
inhabitants, Singapore $4.6$ million). The estimated transmission rate
from Hong Kong data is $\beta \approx 0.75$ and for Singapore $\beta
\approx 0.68$. Both Hong Kong and Singapore's data are
fit with the value $q=0.1$. Hong Kong and Singapore's measure of
contact between diagnosed SARS cases and susceptibles are $l=0.38$ and
$0.40$, respectively (see Figure 4). Even though there is some
heterogeneity in the
parameters for Hong Kong and Singapore, they provide an important
calibration of our model. Their values for $l$ and $\alpha$ are
roughly consistent with each other, indicating that the difference
with Toronto is significant within our model, and pointing to the
joint importance of
rapid diagnosis $\alpha \approx 1$ and good isolation of diagnosed
patients $l \approx 0$ in controlling an outbreak. While there is some
indication in the
data from Hong Kong of a possible slowing of the outbreak, we did not
attempt to analyze the slowing or assess its significance.

\section {Conclusions}
\indent A simple model that can capture the effect of average
infectiousness in a heterogeneous population and the effect of
isolating diagnosed patients has been introduced to explore the role
of
patient isolation and diagnostic rate in controlling a SARS outbreak.
By examining two cases with relatively clean exponential growth curves
for the number of recognized cases, we are able to calibrate a SEIJR
model with parameters $\alpha = 1/3$ (SARS' diagnostic rate) and $l \approx 0.4$ 
(isolation effectiveness). We then use
our SEIJR model to examine
the non-exponential dynamics of the Toronto outbreak. Two features of
the Toronto data, the steep increase in the number of recognized cases
after March 31st and rapid slowing in the growth of new recognized
cases,
robustly constrained the SEIJR model by requiring that $l \approx 0.05$
and $\alpha > 1/3$ days$^{-1}$.

The model is also used to look at the impact of
drastic control measures (isolation). The fitting of
data shows that the initial rates of SARS' growth are
quite similar in most regions leading to estimates of
$R_0$ between $1.1$ and $1.2$ despite the recent identification of
{\it
superspreaders}. Model simulations are fitted to SARS reported data
for the province of Ontario, Hong Kong and Singapore.  Good fits are
obtained
for reasonable values of $\alpha$, the rate of identification of SARS
infections; ``reasonable'' values of the control parameters $l$ (a
measure of isolation); possible values of $p$, a
{\it crude} measure of reduced susceptibility (due to genetic factors,
age or reduced contact rates); $q$ a {\it crude} measure of the
relative
degree of infectiousness of asymptomatic individuals; possible values
of $\rho$ a measure of initial levels of population
heterogeneity; and, reasonable values of N the {\it effective}
population size.  It is worth noticing that for values of $N$ larger
than $100,000$ the predictions (proportion of cases at the end of the
outbreak, etc.) are {\it roughly} the same.  The introduction of
behavioral changes that follow the identification of the first case
(reduce values of $l$ at the time of the identification and moving
aggressively to identify cases of SARS by increasing $1/\alpha$)
result in a dramatic reduction in the total number of cases and on
mortality in Toronto.
Given the fact that SARS appears to kill between three and seven
percent of infected (diagnosed) cases (\cite{BBC1}), it seems quite
appropriate to
isolate diagnosed people. Although we do not examine the effect of
quarantine by varying $q$, it seems intuitive that quarantining
those who came into close contact
with positively diagnosed individuals will reduce the total number of
cases.\\

Model results and simple estimates suggest that {\it local}
outbreaks may follow similar patterns. Furthermore, the use of
relative extreme isolation measures in conjunction with rapid
diagnosis has strong impact on the
local dynamics (Toronto's situation). However,
if SARS has shown us anything it is that
``undetected'' and ``unchecked'' local disease dynamics can rapidly
become a global issue.

The research on this article used the latest data available (April
$18$ for Canada and April $21$ for Hong Kong and Singapore). Recent
disclosures \cite{WP13} reaffirm the importance of carrying out the
analysis excluding data from China. We have redone the analysis 
including the data collected up to April $25$ and, our conclusions, 
remain the
same. Current data seem to support higher values for SARS induced
mortality rates \cite{BBC1}. However, our model is \textit{most 
sensitive} to
the parameters $l$ (effectiveness of isolation) and
($\alpha$) diagnostic rate. It is not as sensitive to
changes in $\delta$. In fact, the consideration of a
$7\%$ mortality  ($\delta \approx 0.01$) rather than $4\%$
reduces the number of cases by about $12\%$. In Toronto, we have
estimated $612$ diagnosed cases with ($l=0.05$ and $\alpha = 1/3$ 
after March
26th). Perfect isolation after March 26th, ($l=0.00$) reduces this 
number to
$396$ diagnosed cases. The assumption of homogenous mixing implies 
that
our model is likely to overestimate the size of the outbreak. Hence,
the situation in Toronto seems to support the view that this {\it
outbreak} is being contained. Obviously, the case of the crude model
(by design) cannot handle high levels of variability (an stochastic
model would be desirable). This possibility is tested (as it is often
done in deterministic models) by looking at the sensitivity of the
model to parameters ($\alpha$ and $l$ being the most critical). Such
sensitivity analyses can also help ``estimate'' the variability in
$R_0$. 

\section{Acknowledgments}

We thank Penny J. Hitchcock, Norman L. Johnson,  Krastan B. Blagoev,
and the T-11 epidemiology discussion group at Los Alamos National
Laboratory and Hans Frauenfelder for enhancing our ability to carry
out this research. We also thank Fred Brauer, Simon Levin, James
Watmough (who reconfirmed our value of $R_0$ by the method
\cite{Diekmann1}), Carlos W. Castillo-Garsow, and Miriam Nuno for
their recent comments. This research has been supported through the 
Center
for Nonlinear Studies
at Los Alamos National Lab under Department of Energy contract
W-7405-ENG-36 and partially supported by NSF, NSA and Sloan Foundation
grants to Carlos Castillo-Chavez. During the final stages of
preparation, it came to our attention that Prof. Roy Anderson is
examining similar questions about SARS' outbreak dynamics.

\vfill

\section{Tables}

\noindent Table 1. Parameter definitions and values that fit the
cumulative number of cases in class $J$ (``diagnosed'') for Hong Kong.
These parameters are used
to compute the basic reproductive number $R_0$.\\
\begin {tabular}{||c|l|c||}
\hline
Parameter & Definition & Value\\
\hline
$\beta$ & Transmission rate per day & $.75$\\
$q$ & relative measure of infectiousness for the asymptomatic class
$E$ & $0.1$\\
$l$ & relative measure of reduced risk among diagnosed SARS cases &
$0.38$\\
$p$ & reduction in risk of SARS infection for class $S_2$ & $0.1$\\
$k$ & rate of progression to the infectious state per day &
$\frac{1}{3}$ \\
$\alpha$ & rate of progression from infective to diagnosed per day &
$\frac{1}{3}$\\
$\gamma_1$ & rate at which individuals in the infectious class
recover per day & $\frac{1}{8}$\\
$\gamma_2$ & rate at which diagnosed individuals recover per day &
$\frac{1}{5}$\\
$\delta$ & SARS-induced mortality per day & $0.006$\\
$\rho$ & Initial proportion of the popualtion at higher risk of SARS
infection & 0.4 \\
\hline
\end{tabular}
\\
\vfill

\noindent Table 2. Estimated starting times of the SARS outbreak.\\
\begin {tabular}{||l|l||}
\hline
     Country & Estimated start of the outbreak \\
\hline
     Canada & February 1st \\
     Hong Kong & November 20th \\
     Singapore &  December 6th \\
     World data & November 5th\\
\hline
\end{tabular}
\\

\vfill
\noindent Table 3. Long-time model results for Ontario, Canada,
assuming various changes in behavior on March 26th, 2003.\\
\begin {tabular}{||l|c|c|c||}
\hline
     $l$ & $\alpha$ & Infected with SARS & Diagnosed with SARS\\
\hline
     $0.05$ & $1/3$   & $0.0077\%$ & $0.0055\%$\\
     $0.3$ & $1/3$   & $18\%$ & $13\%$\\
     $0.05$ & $1/6$   & $21\%$ & $13\%$\\
\hline
\end{tabular}
\\

\section{Figures}

\begin{figure}[h*]
   \begin{center}
   \scalebox{0.6}{\includegraphics{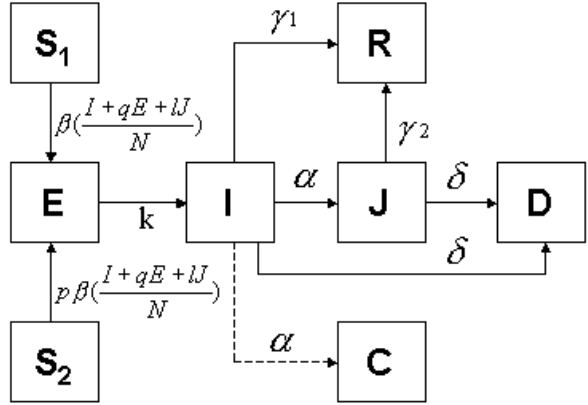}}
   \end{center}
   \caption{A schematic representation of the flow of individuals
   between the different classes. The model considers two distinct
susceptible
classes: $S_1$, the most susceptible, and $S_2$.
$\beta\frac{I+qE+lJ}{N}$ is
the transmission rate to $S_1$ from $E$, $I$ and $J$. $p$ is a measure
   of reduced
susceptibility to SARS in class $S_2$.  $E$ is the class composed of
asymptomatic, possibly infectious
individuals. The
class $I$ denotes infected, symptomatic, infectious, and undiagnosed
individuals.  $I$-individuals move into the diagnosed class $J$ at the
rate $\alpha$.  Individuals recover from class $I$ at the rate
$\gamma_1$ and $\gamma_2$ from the $J$ class. The rate $\delta$
is SARS'
disease-induced mortality. The classes $R$ and $D$ are included to
keep track of the cumulative number of diagnosed, recovered and dead
individuals,
respectively. The quantity $C$ is for comparison with epidemiological
   statistics; it tracks the
total number of diagnosed individuals. }
\label{myfig0}
\end {figure}
\vfill

\begin{figure}[h*]
       \begin{center}


\scalebox{0.6}{\includegraphics{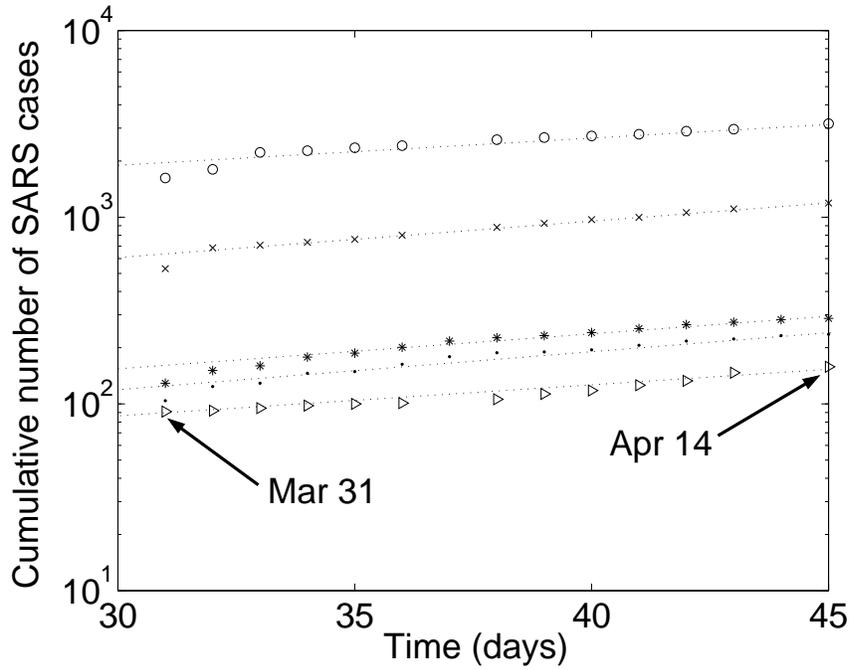}}

\end{center}
      \caption{The cumulative number of SARS cases from March $31$ to April
        $14$ (lin-log scale) for the World (top row), Hong Kong(second
        row), Ontario, Canada (fourth row), all of Canada (third row)
        and Singapore (bottom row). The data were
        obtained from WHO \cite{WHO} except for the Canadian data
        which are from the Canadian Ministry of Health \cite{CMH}.
        The Ontario data includes suspected and probable cases
        since March $31$. This inclusion explains the jump in the
        data for Ontario on March 31st. The rates of growth of the 
SARS
        outbreak (computed
        using data from March $31$ to April $14$) are: 0.041 (world),
        0.050 (Hong Kong), 0.037 (Singapore), 0.054 (Canada) and 0.054
        (Ontario).}
\label{myfig00}
\end {figure}

\vfill
\begin{figure}[h*]
       \begin{center}
       \scalebox{0.6}{\includegraphics{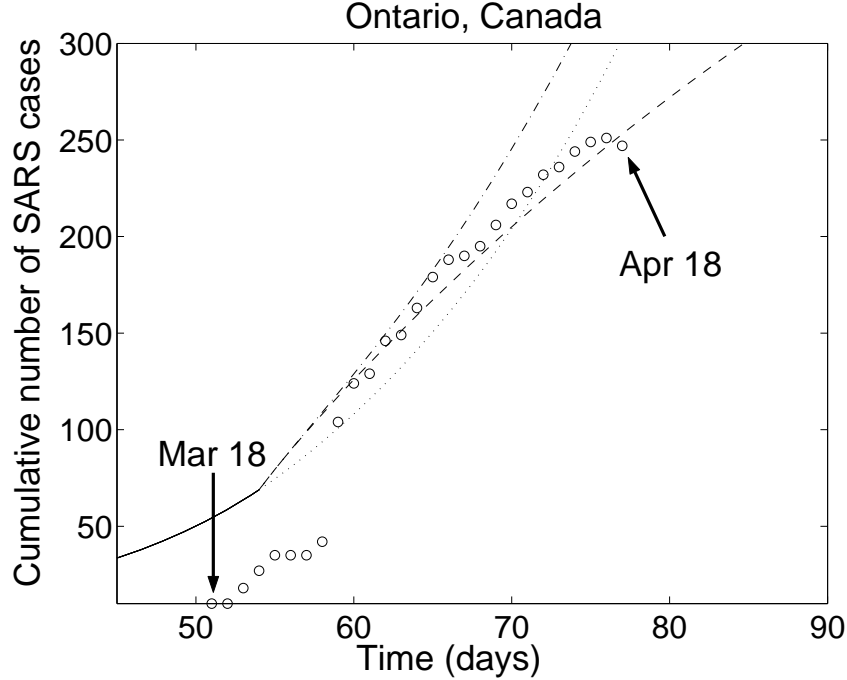}}
       \end{center}
      \caption{The circles are the cumulative number of suspected
        or probable SARS cases in Ontario beginning on day 61 (March
        31st, the day of the jump) and the number of probable cases up
        until day 60. The data prior to day 61 only
        bound the model from below. The lines are the cumulative
        number of ``diagnosed'' cases $C$ from the SEIJR model ($C$ is
        the running sum of all diagnosed cases $J$). The
        fit to the data is given by a change in the values of $\alpha$
and
        $l$ on March 26th. Prior to March 26th, $\alpha=1/6$,
        $l=0.76$. Because the model is poorly constrained prior to day
        61, the real purpose of this part of the model
        is to generate sufficiently large classes of $E$ and $I$
        relative to $J$ on March 26th to give the fast increase in $C$
from
        day 61 to day 67.
        After March 26th, three scenarios are shown. The fit to the
        data is given by $\alpha=1/3$, $l=0.05$ (rapid diagnosis and
        effective isolation of diagnosed cases, dashed line). The
        second curve is given by
        $\alpha=1/6$, $l=0.05$ (slow diagnosis and effective 
isolation,
        dotted line)
        and the third curve by $\alpha=1/3$, $l=0.3$ (rapid diagnosis
        with improved but imperfect isolation, dash-dot line).
        An index case is assumed on February
        $1^{\mathrm{st}}$. The transmission
        rate $\beta$ is computed using the estimated rate of growth
        ($r=0.0543$) for the Ontario data as described in the text.}
\label{myfig01}
\end {figure}

\vfill

\begin{figure}[h*]
       \begin{center}

\scalebox{0.6}{\includegraphics{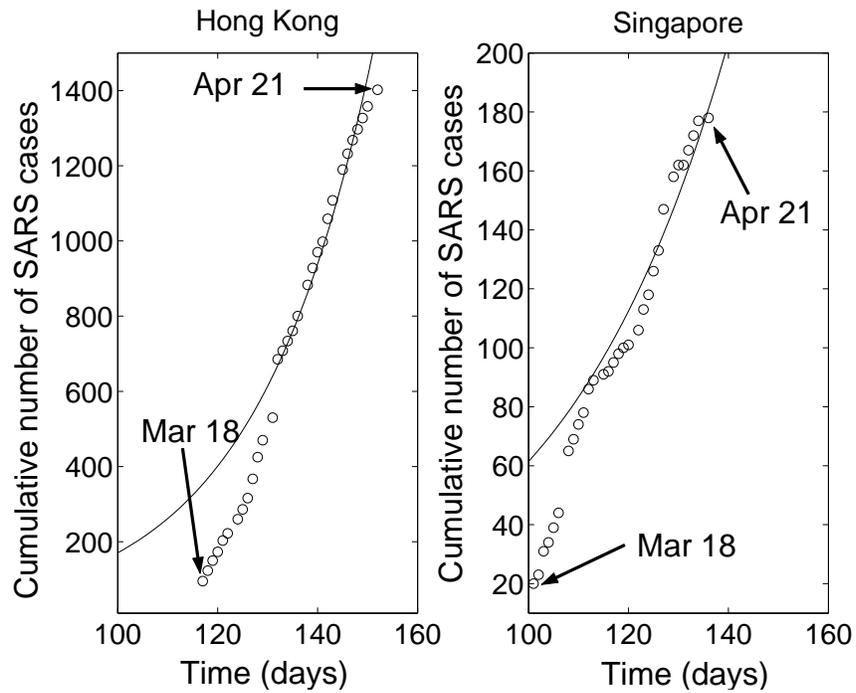}}
       \end{center}
      \caption{Cumulative number of SARS cases in Hong Kong and
      Singapore as a function of time (SEIJR model) with $l=0.38$ 
(Hong
      Kong) and $l=0.40$ (Singapore). Singapore has $\beta = 0.68$,
      all other parameter are from Table 1. The data is fitted
      starting March 31 (see Figure \ref{myfig00}) because of the 
jump in
      reporting on March 30th.}
\label{myfig02}
\end {figure}

\end{document}